\documentstyle[12pt]{article}

\begin{document}
\centerline {\Large {\bf Quest for deconfined quark phase:}}
\centerline {\Large {\bf  A new avenue}}
\bigskip
\centerline { V. S. Uma Maheswari }
\smallskip
\centerline {\bf {\it Variable Energy Cyclotron Centre,}}
\centerline {\bf {\it 1/AF Bidhan Nagar, Calcutta - 700 064, India.}}
\medskip
\centerline {and}
\medskip
\centerline {L. Satpathy}
\smallskip
\centerline {\bf {\it Institute of Physics, Bhubaneswar -751 005, India.}}
\bigskip
\begin{abstract}
{ Stability of droplets of $ud$ matter, {\it viz} udilets, is studied
using non-relativistic and relativistic bag models. It is found
that there is a strong possibility for
forming hypercharged metastable udilets
with charge fraction $Z/A \ge 1$ in semicentral relativistic heavy ion
collisions. It is then speculated that protons hadronised from these
hypercharged udilets may be contributing towards the directed
sideward flow observed recently in experiment.
This exotic state with large charge to mass ratio presents an
unique possibility for
the spontaneous emission of positrons due to the `sparking' of the
vacuum.}
\end{abstract}
\newpage

        The principle goal of present day relativistic heavy ion
collision experiments is to find new states of matter such as
quark gluon plasma\cite{js}. Such pursuits are based on the
 viewpoint that
there exists a transition between nuclear and quark matter[2-12].
Logically, one would expect the transition to take place when
the nuclear matter is compressed to a density at which nucleons
overlap to a significant extent. This density is roughly given by
$\rho_N \sim 3/(4\pi r_N^3)$, where $r_N$ is the nucleon size.
Taking $r_N \sim 0.8 -1.0\ fm$,
we get $ \rho_N \sim 0.24-0.40\ fm^{-3}$.
Consequently, it has been conjectured\cite{sa,ef,hs,cg}
 that a deconfined quark phase(DQP)
 could be produced in heavy-ion collisions(HIC).

Over the years, considerable amount of
work has been done to find
`unambiguous' signatures of DQP\cite{kk}.
In this regard, probable formation
and detection of small lumps of strange quark matter in relativistic
HIC is being promoted as an unambiguous signature\cite{cg}. This
speculation is based on the fact that in ideal baryon rich quark
matter in thermodynamical equilibrium , the production of $s{\bar s}$
pairs is enhanced
compared to the production of light quarks$(u\ \& \ d)$ as
soon as the Fermi energy of the already present $u$ and $d$ quarks  is
higher than twice the strange quark mass. If the temperature and density
attained in a
semicentral/peripheral collision is such that the production of
antiparticles and strange quarks
are suppressed, then the chances are that
a blob of DQP consisting of only $u$ and $d$ quarks, namely an udilet,
will be formed.
The probability of forming an udilet decreases as the collision
becomes either very central or very peripheral.
It would be interesting to understand how does such a state show
itself, and if there is any specific signal,
quite different from those characteristic
of very central collisions. In this work, we mainly focus upon
these aspects and investigate them using
a non-relativistic model(NRM)\cite{ls} as well
as the usual relativistic bag model(RBM)\cite{jm}.

Recently, we have developed\cite{ls}
a consistent mass formula for multiquark
droplets based on a non-relativistic framework. This formula is quite
successful in describing the properties of strangelets, and also is
found to be fully consistent with relativistic studies.
Omitting the details, we give here only the final expression.
Consider a system of $N_u$ and $N_d$ number of $u$ and $d$ quarks
respectively, with mass $m$, confined in a spherical box of size $R$.
For a given baryon number $A$, we define an asymmetry parameter,
$\delta = (N_d-N_u)/N_q$, where $N_q=N_u+N_d=3A$. Then the quark
flavor densities are defined as $\rho_u =0.5\rho_{oq}(1-\delta)$
and $\rho_d=0.5\rho_{oq}(1+\delta)$, where $\rho_{oq}$ is the
equilibrium total quark number density. We can then calculate, for a given
$\delta$, the total mass energy
of an udilet using the expression\cite{ls},
\begin{eqnarray}
M_{\rm NR}(A)
&=& {5\over 3}\ {\hbar^2\over 2m}\
\Large [ {3\pi^{4/3}\over 5} \Large ]
\left ( \rho_u^{5/3} + \rho_d^{5/3} \right ) \ {4\pi R^3\over 3}
\nonumber \\
&\quad& + {5\over 3}\ {\hbar^2\over 2m}\
\Large [ {3\pi^{5/3}\over 16} \Large ]
\left ( \rho_u^{4/3} + \rho_d^{4/3} \right ) \ {4\pi R^2}
\nonumber \\
 &\quad& + {5\over 3}\ {\hbar^2\over 2m}\
 \Large [ {9\pi^{2}\over 128} - {1\over 3} \Large ]
\left ( \rho_u + \rho_d \right ) \ {8\pi R}
\nonumber \\
&\quad & + {4\over 3}\ \left ( {0.6Z^2e^2\over R} \right ) + 3mA,
\end{eqnarray}
given in terms of the equilibrium baryon number density
$\rho_{ob}$(=$\rho_{oq}/3$) and the total charge $Z=0.5A(1-3\delta)$.
Here, we take the constituent quark mass $m$ to be 300 MeV\cite{ls}.
We also give the corresponding expression obtained in the case of the
usual relativistic bag model\cite{jm}, where
the $u$ and $d$ quarks are massless.
\begin{eqnarray}
M_{\rm RLV}(A)
&=& {4\over 3}\ \Large [ {3\pi^{2/3}\over 4} \Large ]
\left ( \rho_u^{4/3} + \rho_d^{4/3} \right ) \ {4\pi R^3\over 3}
\nonumber \\
&\quad& + {4\over 3}\ \Large [ {\pi^{-2/3}\over 8} \Large ]
\left ( \rho_u^{2/3} + \rho_d^{2/3} \right ) \ {8\pi R}
\nonumber \\
&\quad & + {4\over 3}\ \left ( {0.6Z^2e^2\over R} \right ).
\end{eqnarray}
It may be noted that unlike in the non-relativistic model, the surface
energy coefficient for an udilet is zero in the present case.
Here, we have omitted the
lowest order quark-quark interaction with the
supposition that it can effectively be absorbed into the
bag parameter, or equivalently into the saturation
density $\rho_{ob}$ as has been done by Berger and Jaffe\cite{bj}.

We then calculate the mass per baryon of an udilet at the
 equilibrium density
$\rho_{ob}$ for various values of charge fraction $f_Z=Z/A$.
We have taken $\rho_{ob}=0.25\ fm^{-3}$ in the case of NRM.
And in the case of RBM, the value of $\rho_{ob}$ is taken to be
0.40$fm^{-3}$ so that
$M_{\rm RLV}(\rho_{ob} )/A > M_N$, where $M_N$=939 MeV.
Results so-obtained using Eqs.(1) and (2) are shown in Fig.1 for a
representative value of $A$.
It is seen that the mass per baryon
shows a well defined minimum at
$f_Z^{min} \sim 0.4$
for $A=300$ in both the models.
Further, we find in both the models,
$f_Z^{min}$ increases as $A$ is decreased
and/or $\rho_{ob}$ is increased.
It is satisfying to find that our non-relativistic model
qualitatively agrees with
the relativistic bag model.

Let us now suppose that an udilet with $f_Z \simeq 0.4$ and
$A \simeq 300$ is formed in a
semicentral/peripheral HIC at relativistic energies.
The two important channels for the decay of the
udilet are fission and strong nucleon emission.
Firstly, let us investigate whether the udilet undergoes fission.
For fission to be energetically favourable,
we must have the mass energy difference
$E_f = M(A,Z)-M_1(A_1,Z_1)-M_2(A_2,Z_2) \ge 0$. Considering
symmetric fission and a liquid drop model like expansion for the
mass energy[Eq.(13) of Ref.\cite{ls}], one can rewrite the criterion
$E_f \ge 0$ as $Z^2/A \ge 0.26a_s/0.37a_C$, where $a_s$ and $a_C$ are
respectively the surface and Coulomb energy coefficients.
In the case of RBM, where $a_s \sim 0$, the criterion $E_f \ge 0$ can
be expressed in terms of the curvature energy coefficient $a_{\rm cv}$
as $Z^2/A^{2/3} \ge 0.59a_{\rm cv}/0.37 a_C$. Further, we are aware
from the nuclear fission studies that eventhough $E_f \ge 0$,
instantaneous fission does not always happen.
An additional criterion for such a decay process is that
$E_f/E_b \ge 1$, where
$E_b = Z_1Z_2e^2/[r_0 (A_1^{1/3}+A_2^{1/3})]$ is the Coulomb barrier
energy and $r_0 = {\Large ( 3/(4\pi \rho_{ob}) \Large )}^{1/3}$.
In Fig.(2), we have plotted $E_f/E_b$ calculated using
Eqs.(1) and (2) as a function
of $Z/A$, for $A=300$ and $A=40$.
In the NRM, for $A=40$, fission is not allowed. For $A=300$,
$E_f/E_b$ being negative upto $f_Z \simeq 1.6$, fission is
disallowed and over the range $f_Z \sim 1.6-2.0$, though
$E_f/E_b$ is slightly positive, instantaneous fission is
disallowed.
But, in the RBM, udilets with large value of $A$ are susceptible
towards fission, which may be due to the vanishing of $a_s$.
For example, $E_f/E_b \sim 1$ at $f_Z \sim 0.7$, for $A=300$; and
at $f_Z \sim 1.0$, for $A=200$.
Further, for a given $f_Z$, the ratio $E_f/E_b$ becomes
more negative as $A$ decreases and/or $\rho_{ob}$ increases.
However, in the case of relativistic bag model as the
finite-size coefficients such as curvature $a_{\rm cv}$ and
Coulomb $a_{C}$ are proportional to $\rho_{ob}^{1/3}$, the
ratio $E_f/E_b$ is independent of $\rho_{ob}$.
Since, an udilet formed in a HIC is most likely to have
$A\le 300$ and $f_Z \simeq 0.3-0.5$ and fission
is disallowed for these values of $A$ and $f_Z$ in both
the models, the predominant channel for decay is the
strong nucleon emission.

Among the baryons like neutron, proton and $\Delta$'s, the neutron
emission is the most favourable one
because of low mass and non-existence
of Coulomb barrier.
As an udilet with $f_Z \simeq 0.4$
and $A \simeq 300$
emits neutrons, it is driven towards higher values of $f_Z$.
This process is energetically allowed as long as
$Q_n=M(A,Z)-M(A-1,Z)-M_N \ge 0$.
We find that once the charge fraction $f_Z$ reaches a value of
about 1.2 with $A \simeq 102$ and
1.9 with $A\simeq 64$ in the cases of
RBM and NRM respectively, the neutron emission stops.
It is also seen that the limiting values of $f_Z$ increases
as $A$ decreases and/or $\rho_{ob}$ increases.
It maybe noted that the limiting value of $f_Z$ obtained in
the case of RBM is lower than the one obtained in the
case of NRM.
At this stage of hyper-chargeness, the system is unstable
against proton, $\Delta^+$ and $\Delta^{++}$ decays.
Here, the proton emission is the most favourable one,
and is allowed as long as the corresponding $Q-$value,
$Q_p \ge 0$.
We find in the case of NRM, the proton emission
continues upto $f_Z \sim 2$ with $A \sim 58$, where the
entire udilet is composed of only $u$ quarks.
This non-relativistic
system is highly unstable against strong $\Delta^{++}$ decay, and
thus it finally disintegrates.

In the case of RBM, the proton emission
continues upto $f_Z \sim 1.9$ with $A \sim 20$. At this stage, the
system is unstable against strong $\Delta^{++}$ decay.
However, it must be noted that for such small baryon numbers, shell
effects are important\cite{gj} and
shall decide the stability of the udilet.
For this reason, we recalculated the Q-value,
$Q_{\Delta^{++}}=M(A,Z)-M(A-1,Z-2)-M_{\Delta^{++}}$, using the
shell model calculations\cite{gj,dv}.
The total mass energy $M(A,Z)$ of an udilet
at the saturation point is obtained using the expression,
\begin{equation}
M(A,Z) = {4\over 3R}\left [ \sum_{i=1}^{N_u} \ g_i \omega_i +
\sum_{i=1}^{N_d} \ g_i \omega_i + 0.6Z^2e^2 \right ],
\end{equation}
where $R = r_0 A^{1/3}$ and $g_i$ is the degeneracy factor.
It may be noted that the values of $A$ and $f_Z$ of the
udilet remaining after
the emission of neutrons and protons
depend upon the characteristics$(A,f_Z,\rho_{ob})$ acquired
at its birth. For example, if the initial value of $A$ is 200 instead
of 300, then the limiting values at the end of neutron and
proton emissions are $A\sim 14$ and $f_Z \sim 2$.
Therefore, it may be appropriate to study the dependence
of $Q_{{\Delta}^{++}}$ on $A$, $f_Z$ and $\rho_{ob}$.

To do so, we calculated $Q_{\Delta^{++}}$
and the Coulomb barrier energy
$V_{CB}=2(Z-2)e^2{(A-1)}^{-1/3}/r_0$ for various values of $A$,
keeping fixed $\rho_{ob}$ and $f_Z$.
Results are displayed in Figs.(3) and (4) for two values of $f_Z$
and three values of $\rho_{ob}$.
It is interesting to see that certain udilets are absolutely
stable against $\Delta^{++}$ decay, i.e. $Q_{\Delta^{++}} <0$.
For a given $A$, the stability increases as $f_Z$ and $\rho_{ob}$
decreases. Under these conditions, udilets can decay only by
converting some of the $u$ quarks into $d$ quarks
through weak leptonic decay process.
Hence, this study predicts quite a good possibility for the
formation of metastable udilets of relatively long decay lifetime,
which is likely to be observed in experiments.
The characteristic signature of such udilets is
their large charge to mass ratio, $Z/A \ge 1$,
 which is in sharp contrast to the low value $Z/A <<1$
found in case
of strangelets.

Thus, from both the relativistic and non-relativistic studies,
 an optimistic scenario about DQP in form of udilets with
large charge to mass ratio as characteristic signature emerges.
While in former case metastable udilets with relatively large
lifetime are formed, in the latter case they disintegrate
by proton, $\Delta^{+}$ and $\Delta^{++}$ decay processes.
The predominance of proton in this phenomenon is quite evident.
This is suggestive of the interesting possibility that the
`direct' protons as well as those from
$\Delta^{++} \longrightarrow p+\pi^+$ decays, originating from
the udilets, may be
contributing toward the directed sidewards flow observed recently
\cite{jb}. In this context, the observations that protons are the
major carriers of flow and the magnitude of the flow decreases as the
collision becomes very central are quite inspiring to us.
Further, in the present study we have shown that the udilets
formed after
neutron emissions has $f_Z \simeq 1.2-1.9$, depending
upon the model. This presents a novel possibility
of a system where charge number substantially exceeds
the mass number. This then provides a highly favourable
situation for reaching the critical value for $Z\ (\simeq 137)$
for the occurence of the
spontaneous emission of positrons due to the `sparking' of the vacuum
\cite{jg}.
Finally, we would like to comment that for the aforesaid possibilities
to be realised, the initial parameters of an udilet, i.e. $A$, $f_Z $ and
$\rho_{ob} $, should be such that the charge fraction acheived at the
end of neutron emissions is greater than one.

In conclusion, stability of udilets is studied
using non-relativistic and relativistic bag models. It is found
that there is a strong possibility for
forming hypercharged metastable udilets
with charge fraction $Z/A \ge 1$ in semicentral relativistic heavy ion
collisions.
This large value of charge fraction can serve as an unique signature for
DQP. This is in sharp contrast to the strangelets which are
characterised by small charge fraction.
It is then speculated that protons hadronised from these
hypercharged udilets may be contributing towards the directed
sideward flow observed recently in experiments.
An exciting novel possibility of forming a system having
large charge number with relatively small baryon number
raises new hope of the spontaneous emission of positrons
due to the `sparking' of the vacuum.

\noindent We are grateful to Dr. S.K. Gupta and Prof. C.V.K. Baba for
very useful discussions. One of us  would like to
acknowledge Institute of Physics, Bhubaneswar, for hospitality,
where part of this work was completed.

\newpage

\centerline {\bf FIGURE CAPTIONS }
\vskip 1.0 true cm

\noindent {\bf FIG.1 \ : \ }
Total mass per baryon of an udilet calculated using Eqs.(1)[NRM]
 and (2)[RBM] are shown
as a function of the charge fraction $Z/A$ for baryon number $A=300$.
\vskip 1.0 true cm

\noindent {\bf FIG.2\ : \ }
Ratio of the fission energy $E_f$ to the Coulomb barrier energy $E_b$
calculated using Eqs.(1)[NRM] and (2)[RBM]
are shown as function of the charge fraction $Z/A$ for two values of
baryon number $A$.
\vskip 1.0 true cm

\noindent {\bf FIG.3\ :\ }
Values of $Q_{\Delta^{++}}/V_{CB}$ obtained from shell model
calculations[Eq.(3)]
are shown as a function of baryon
number $A$ for three values of saturation baryon number density $\rho_{ob}$,
keeping fixed the charge fraction $f_Z=Z/A$ at 1.8.
\vskip 1.0 true cm

\noindent {\bf FIG.4\ :\ }
Same as Fig.3, but for $f_Z=1.3$.

\vfill

\vfill
\newpage
\begin{thebibliography}{99}
\bibitem{js} J. Stachel and G.R. Young, Annu.Rev.Nucl.Part.Sci.{\bf 42},
 537 (1992) and references therein.
\bibitem{jc} J.C. Collins and M.J. Perry, Phys.Rev.Lett.{\bf 34},
1353 (1975).
\bibitem{gb} G. Baym and S. Chin, Phys.Lett.{\bf 62B}, 241 (1976).
\bibitem{mk} M. Kislinger and P. Morley, Phys.Lett.{\bf 67B}, 371 (1977).
\bibitem{sa} S.A. Chin and A.K. Kerman, Phys.Rev.Lett.{\bf 43}, 1292 (1979).
\bibitem{ew} E. Witten, Phys.Rev.{\bf D30}, 272 (1984).
\bibitem{ef} E. Farhi and R.L. Jaffe, Phys.Rev. {\bf D30}, 2379 (1984).
\bibitem{bj} M.S. Berger and R.L. Jaffe, Phys.Rev. {\bf C35}, 213 (1987).
\bibitem{fc} F. Curtis Michel, Phys.Rev.Lett.{\bf 60}, 677 (1988).
\bibitem{ca} C. Alcock and A. Olinto, Annu.Rev.Nucl.Part.Sci.
{\bf 38}, 161 (1988).
\bibitem{hs} H. Stocker and W. Greiner, Phys.Rep.{\bf 137}, 277 (1986)
and references therein.
\bibitem{cg} C. Greiner, P. Koch and H. Stocker, Phys.Rev.Lett.{\bf 58},
1825 (1987).
\bibitem{kk} K. Kajantie and L. McLerran, Annu.Rev.Nucl.Part.Sci.
{\bf 37}, 293 (1987) and references therein.
\bibitem{ls} L. Satpathy, P.K. Sahu, and V.S. Uma Maheswari, Phys.Rev.
{\bf D\ 49}, 4642 (1994).
\bibitem{jm} J. Madsen, Phys.Rev.{\bf D\ 50}, 3328 (1994).
\bibitem{gj} E.P. Gilson and R.L. Jaffe, Phys.Rev.Lett.{\bf 71},
332 (1993).
\bibitem{dv} D. Vasak, W. Greiner and L. Neise, Phys.Rev.{\bf C34},
1307 (1986).
\bibitem{jb} J. Barrette et al., Phys.Rev.Lett.{\bf 73}, 2532 (1994).
\bibitem{jg} J.S. Greenberg and W. Greiner, Physics Today, August 1982, p.24.
\end{thebibliography}
\end{document}